# Reviewing the Need for Explainable Artificial Intelligence (xAI)

Julie Gerlings
Copenhagen Business School
jge.digi@cbs.dk

Arisa Shollo
Copenhagen Business School
ash.digi@cbs.dk

Ioanna Constantiou
Copenhagen Business School
ic.digi@cbs.dk

## Abstract

*The diffusion of artificial intelligence (AI) applications in organizations and society has fueled research on explaining AI decisions. The explainable AI (xAI) field is rapidly expanding with numerous ways of extracting information and visualizing the output of AI technologies (e.g. deep neural networks). Yet, we have a limited understanding of how xAI research addresses the need for explainable AI. We conduct a systematic review of xAI literature on the topic and identify four thematic debates central to how xAI addresses the black-box problem. Based on this critical analysis of the xAI scholarship we synthesize the findings into a future research agenda to further the xAI body of knowledge.*

## 1. Introduction

Machine learning (ML) / Artificial Intelligence (AI) technologies including neural networks (NN) variations, are evolving at a rapid pace and thereby expanding their capabilities, resulting in advanced models being used more frequently in decision making processes [1]. As these models are integrated in organizations and daily work, academics and practitioners need to pay more attention to the development process of the models and the interpretation of the outcome [2]–[4]. This is important as vital decisions are increasingly being supported or fully automated by different forms of algorithms that are not fully understood by people, - known as the AI black box explanation problem [5]. If we cannot explain the algorithms, we cannot argue against them, verify them, improve them, or learn from them [6], [7]. Both practitioners and academics call for better understanding of the complex and opaque models by implementing transparent and understandable ML-models. Generating explanations of how the ML-models work and how results are produced, leads to increased trust in machine learning models [5], [8]–[10].

Explainable AI has emerged as a response to the increasing "black box" problem of AI, according to which models and their performance are not understandable by humans. xAI refers to methods and techniques seeking to provide insights into the outcome of a ML-model and present it in qualitative understandable terms or visualizations to the stakeholders of the model [9], [25], [33].

xAI initially started from the computer science community who were building xAI methods providing a technical solution to the problem. With the diffusion of AI applications in business and society xAI has evolved to a broader research domain. From being viewed as the extraction of advanced statistical values for developers to identify information gain and entropy of variables, more advanced frameworks like LIME offer user friendly visualization of local instances for the user to interpret. The emerging adoption of AI adds another layer of complexity to human-computer-interaction (HCI), where xAI has the potential to play a central role in the behavioral and cognitive aspects of AI- assisted decision making. The AI-based systems are not new, they are descendants of expert systems [92] utilizing more advanced technologies (e.g. machine learning algorithms). Yet, they suffer from opacity in terms of their ability to explain how conclusions are reached. Hence, the xAI technology could potentially either clutter or better explain the decisions made or assisted by an algorithm. The critical and systematic analysis of published research constitutes an important vehicle in understanding the current state of xAI scholarship [15]. This analysis identifies the major thematic debates within xAI research and proposes future research directions.

We reviewed xAI literature focusing on purposes, definitions and actions related to xAI. We respond to the call for focusing on metahuman systems - a unique kind of sociotechnical systems - which are blurring the boundaries between the social and the technical in unanticipated ways [11]. We consider xAI as such a system that calls for novel inquiry [11], [16].

Our contribution is twofold. First, we identify four thematic debates central to how xAI research addresses the need for explainable AI. Second, taking a sociotechnical perspective in assessing the debates we identify two future research avenues: a) the need for a stakeholder approach and recognizing that different stakeholders have different explainability needs b) the need for a holistic view on explainability - jointly accounting for the social and the technical aspects, the process and the outcome aspects, as well as the factual and the storytelling aspects of xAI. We argue that to advance theories and practice on xAI, the Information Systems (IS) field is in need of empirical studies that show how different xAI frameworks address the different stakeholder needs.

In the following sections, we first describe the methods used to conduct a systematic literature review. This is followed by our findings, which we structured around four thematic debates in the xAI literature. Finally, we discuss the findings and propose a future research agenda for xAI research.

## 2. Method

We performed a systematic literature review using strategies from Webster and Watson [17], and Jones and Gatrell [18] focused on creating a theme- centered retrospective view of current literature in xAI, generating insights about the current debates and identifying gaps for guiding further research [19]. The review consists of search, selection, analysis, and synthesis processes. Our aim was to provide a critical analysis of the field rather than providing a descriptive overview [18], [20].

### 2.1. Search and Selection Criteria

In order to identify relevant literature for this study and accommodate the interdisciplinary field of xAI, we used a broad collection of scientific databases: ArXiv, Association for Information Systems (AIS), JSTOR, ACM Digital Library, IEEE Xplore Digital Library, SAGE, and Science Direct. These databases were chosen to ensure a diversity in both computer science and social sciences for a better representation of papers. Due to the emerging nature of the field of xAI, ArXiv was included in the selection of databases to allow previews and submission of papers for journal articles and conferences.

We used the terms 'xAI', 'explainable artificial intelligence' or 'explainable AI' to search for articles in the above databases. The reason for not including 'interpretable' is partly due to the shift in terminology in recent years [21], [22], and the notion that 'interpretation' has mostly been focused on technical methods for information extraction and not generating explanations [13]. We searched for articles where the terms appeared in titles, keywords, abstracts, and full text. Due to the novelty of the xAI field and its rapid development we conduct the search for the last 5 years from 2016-2020. This search resulted in a total of 221 articles comprised of journal articles and conference papers.

Next, articles were screened to determine their relevance based on their abstracts and full text. Inclusion criteria were theme-related to the purpose of the literature review. Papers discussing the purpose, need, reason and capabilities of xAI were initially included. Conceptual and technical studies focusing only on technical performance of xAI methods and techniques that did not engage in a discussion of the xAI problem or need were discarded. Thus, when a method was explained, without paying attention to issues such as validity, interpretation of output, or trust, but merely generating transparency in opaque models, the study would be excluded. Along the same lines, discussion and opinion papers advocating for xAI as a way to increase transparency of ML-models without further elaboration were also discarded.

The search strategy included going both backwards and forwards searching in citations to identify prior articles of interest for possible inclusion, and identify key articles with high impact in the field of xAI [21]. After the inclusion of further literature and selection based on the above inclusion and exclusion criteria, the final review included 64 papers.

### 2.2. Literature Analysis and Synthesis

Considering xAI as a metahuman system - a unique kind of sociotechnical systems [11] we examined the articles with a sociotechnical lens in mind [16]. A sociotechnical approach takes a holistic view where relations among people, technology, tasks and organization are sustainable. From previous IS research we know that *"poorly designed sociotechnical systems with inadequate concern with mutual relationships were shown to fail and produce unintended or unwanted outcomes"* [11, p. 8]. Taking this as a departure point we examine the article pool if concerns like a) a holistic view of social and technical aspect are considered in the xAI literature; b) consideration or participation of relevant stakeholders in xAI design, development and use processes.

During the analysis the primary researcher read all the articles thoroughly and identified topics that were discussed across articles. For each paper, key findings were included in a summary and comments on their different approaches to the topic was noted [17] while open coding [90] was applied by highlighting insights that seem relevant to the review's scope i.e. how the papers addressed the need for xAI. In this way, the primary researcher built a set of concepts and insights based on the excerpts supported by the papers. As a result, a concept matrix was made that was enriched and updated as more articles were read. In the next stage, all three researchers engaged in axial coding [90] in order to identify higher order categories by understanding how some concepts relate to other concepts in the matrix. The analysis of the article pool was ongoing and evolving in an inductive manner, as thematic debates emerged under a) motivating the need for xAI, b) the completeness vs interpretability dilemma c) explanation for humans, d) technical frameworks for xAI. As themes took form, some themes cover broader areas than others as their nature is more diffused.

## 3. Findings

In this section, we present the four thematic debates that emerged from analyzing the article pool.

### 3. 1. Motivating the Need for xAI

Exploring the recent literature on xAI and the purpose of the technology, we observe conceptual differences on how xAI is defined.

There are various interpretations of the foundational concepts such as explanation vs. interpretation and their related concepts. Some researchers use both terms interchangeably, while others depict the differences in the two conceptual chunks.

Miller [8] articulates how an **explanation** in social science is regarded as a two-step process consisting of a) the cognitive process, *explanandum* describing the cause for an event whereof a subset of causes is selected as the explanation (*explanas*) and b) the social process of transferring knowledge between explainer and explainee in an interactive manner. Whereas Brandão et al. [23] takes their stance describing a 'good explanation' as an explanation where the explainer understands what the explanation *means* to the person who asks for it, as they stress the need for investigating the social meaning of what it means to others than developers and researchers.

As Brian and Cotton argue, the terms of interpretability and explainability (and the variations hereof) are intertwined and still cluttered in their definitions *"Explanation is closely related to the concept of interpretability: systems are interpretable if their operations can be understood by a human, either through introspection or through a produced explanation."* [24].

Other scholars [9], [22] take a more pragmatic view and argue that 'interpretation' is closer to the development of models and constitutes the opposite of a 'black-box' model, where the search for a direct understanding of the mechanism by which a model works is the aim of interpretable ML.

Others define interpretation as the means to explain or to present in understandable terms to a human [25] and directs the research in terms of how humans interpret information. Liao et al. argue for a more diverse approach to xAI, taking into account the different user needs. They describe xAI as *"…an example, one of the most popular approaches to explain a prediction made by a ML classifier, as dozens of XAI algorithms strive to do, is by listing the features with the highest weights contributing to a model's prediction"* [26] which for developers might be of high value, however not for the average layman. These different definitions reveal the need for further conceptual alignment in the xAI field. The following sections present the main drivers for xAI systems.

#### 3.1.1. Generate Trust, Transparency and Understanding.
Generating trust, is a major driver for xAI and is strongly related to transparency. DARPAs XAI program promotes the need for xAI, as we need to further understand, trust and manage the emerging generation of artificially intelligent machines [27]. Along these lines, lies a great effort and research focused on extracting information from models or build simpler models in the quest of delivering both transparency, understanding and thereby generate trust in the models [8], [22], [28]– [34]. Gilpin et al. [34] argue that*:" … models that are able to summarize the reasons for neural network behavior, gain the trust of users, or produce insights about the causes of their decisions..."* [34]. Along with DARPA, the general increase in ML performance and use has created the search for better understanding of the models to increase trust and thereby an increased use of ML in the industry [35], [36]. Moreover Miller [8] argues that two complementary approaches will generate more transparent, interpretable, and explainable systems to thereby make us more equipped to understand and trust the models [8]: 1) Interpretability and Explainability understood as how well a human understand an explanation in a given context and 2) An explanation of a prediction (decisions) to people (target audience). Most technical xAI approaches aim at extracting information from a model (it may be e.g. neural networks or random forest) such as feature importance, relative importance scores, sensitivity analysis, rule extractions or other methods to generate greater transparency. These xAI approaches and frameworks mainly work from the perception of transparency may improve understanding and thereby increase trust – or the opposite that 'black-box' models are not to be trusted [9], [31], [33], [37].

Few papers include socio-technical aspects in the technical models presented. However few address the obstacles in stakeholders understanding the output which includes a consideration of the output as a socio-technical aspect of an explanation, a HCI- dilemma and addressing the risks in producing explanations for developers, created by developers (inmates running the asylum) [38], [39]. For example Zang and Zhu [40] presents a graphic logic (or symbolic logic) to ease the understanding of Convolutional Neural Networks (CNN), instead of only information extraction; While Mueller et al. [29] visualizes pixels used to determine a husky from a wolf to test participants through LIME. In this way, they address the need for human understanding by testing their explanation on participants evaluating whether they trust the algorithm [33]. Furthermore, the literature emphasized that the xAI frameworks generating explanations are built by developers or technical people focusing on the computational problems in extracting data, which will not necessarily solve the issue of trust [8], [23], [38], [41], [42].

However, many conceptual papers call for interdisciplinary work and discuss the need for more focus on human understanding or interpretability, and not only transparency [9], [21], [25], [26], [34], [43].

### 3.1.2. Ensure Compliance, follow Regulations and GDPR Laws.
One of the many reactions to new regulations and GDPR laws is the call for xAI to provide explanations not only to the users, but society as a whole [22]. This, along with other regulations, makes it urgent for practitioners and industries to ramp up the investment in explaining opaque models [29], [34]. The GDPR regulation and 'the right to an explanation' has caused great stir within both research and the industry, directing them towards xAI - as a possible solution for being compliant [44]–[46]. Moreover, some researchers argue for regulation of xAI itself, or the possibility of setting a standard or quality measure to ensure a responsible use of xAI and avoid building persuasive models instead of explainable ones [47], [48]. The fallacies of building persuasive explanations are very well described in Gosiewska and Biecek's [49] examples of how additive models can cause misleading guidance on instance-level explanations which is backed by Rudin [50] who argues against the recent trend of building (additive) explainable post-hoc "misleading" explanations.

### 3.1.3. For the Sake of Social Responsibility, Fairness and Risk Avoidance.
Especially, within healthcare, clinical and justice work, risks and responsibility are a major concern, as they are potentially dealing with human lives and not merely cost-benefit analyses [9], [51], [52]. Risk avoidance occurs as responsibility is assigned to the individual professional. Hence, developing mental models for expert (e.g. clinical) reasoning to develop better understanding of the reasoning behind deep neural networks and opaque models [53]–[55]. Moreover, recent events of discrimination and recidivism in opaque models have fueled the debate on ensuring fairness in model performance and deeper understanding of how they are built. Cases of minorities in hiring processes [56], recidivism in the COMPAS system and general fairness [48] have added to the surge in the xAI literature [6], [22], [34], [47], [57].

### 3.1.4. Generate Accountable, Reliable and Sound Models for Justification.
A theme that has caused great attraction towards xAI is the possibility to ensure fairness and unbiased models by auditing them or create proof of their rightfulness. Adadi and Berrada [21] follow this approach and argue that xAI provide the required results for auditing the algorithms and generates a provable way for defending algorithmic decisions as being fair and ethical. Hence, generating algorithms that are not only fair and socially responsible, but also accountable and able to justify their output is another aspect motivating the need for xAI. Amongst others, Abdul et al. [51] describes the complications herein and present one of the more popular approaches to generate xAI, by building counterfactual explanations which are based on the notion of causality and not just correlation. When looking at the specific xAI methods, Doshi-Velez and Kim [25] argue that global explanations of entire models or groups are more appropriate for scientific understanding or bias detection in models, whereas local explanations are better suited for a justification of a specific decision. Moreover, Liao and Anderson [58] present methods for generating argumentation-based justifications and explanations, based on formal argumentation which provide natural arguments for better reasoning of the models. Lastly, Ananny and Crawford [59] bring forward a discussion of transparency not being sufficient to govern and hold algorithms accountable. They present different pitfalls in the ideal transparent model and its limitations. They claim that transparency does not necessarily build trust as different stakeholders trust systems differently, depending on their confidence upon when and what information is disclosed, and how accurate and relevant it is perceived to be [59].

### 3.1.5. Minimize Biases and Misinterpretation in Model Performance and Interpretation.
Biases in models and their performance have shown to be an important driver for xAI, as media coverage of models performing sub-par to humans in e.g. filtering out appropriate candidates in hiring processes [60] or failing at recognizing people of color [61]. Especially when dealing with neural network learning patterns from training data, biased training data becomes an issue that impacts the validity of the model output [62]. For these reasons, this sub-theme is also tightly linked to the two previous sub-themes, as biased or discriminatory models and their results are to no purpose if not minacious in their implementation. Besides biased training data [22], [63], variable selection [64], [62] and representation [32], [33] , our own cognitive bias can furthermore hinder our interpretation of the visualized output from a model as we tend to oversimplify the information [52]. Our cognitive biases are argued to be mitigated by xAI frameworks of various kinds, e.g. by Wang et al. [65] who argue for pursuing reasoning theories and Arrieta et

Page 1287

al. [22], who argue that human cognitive capabilities favors visual presentation of data.

**3.1.6. Being Able to Validate Models and Validate Explanations Generated by xAI.** In response to biased models and sub-par performance to humans, researchers have found four types of evaluation methods for deep neural networks (DNN) measured in 1) completeness compared to the original model, 2) completeness as measured on substitute tasks, 3) the ability to detect biases within a model, 4) human evaluations [34]. Evaluating xAI is not only a question of precision and feature extraction, as the user of the output might not be able to understand the model output. Others present a thorough taxonomy for evaluation of interpretability, where the costliest is the application-grounded approach which entails testing of an implemented explanation and letting end-users test it [25]. In this case the explanation is evaluated based on identification of errors, new facts or less discrimination, to the baseline of human-performance [25]. Furthermore, they present a human-grounded evaluation for testing more general notions of quality such as which type of explanation is best under time constraints. The last evaluation approach is functionally-grounded and fits the evaluation of interpretability in already evaluated models or immature model testing. This approach requires no human interaction but rather measures optimization or quality [25].

## 3.2. Completeness vs. Interpretability Dilemma

From the debate of evaluating xAI, a debate of whether we are able to make good explanations emerges. The tradeoff between a correct (complete) and a good (interpretable) explanation, is discussed by Kim [66] where a pragmatic approach is needed if the user (audience) should be able to understand the explanation. Furthermore, Kim [66] continues to present the notion of a 'grasp-ability' test to ensure the audience can use an explanation that is not necessarily perfectly transparent or rigorous, but graspable to the audience. However, researchers have argued that the need for explainability stems from incompleteness producing different biases and argue that the nature of the user's expertise will influence the level of sophistication the explanation can contain [25].

Many other researchers argue for the same tradeoff between completeness (the goal of describing the operation of a system in an accurate way) and interpretability (here, the goal of interpretability is to describe the internals of a system in a way that is understandable to humans), as interpretability alone is insufficient [34]. Hence, we should be cautious with this tradeoff as humans have a strong specific bias towards simple descriptions which can lead researchers to create persuasive systems rather than transparent systems. Humans lose trust in the explanation when soundness is low [21], [34].

## 3.3. Human Explanations

In general, the literature agrees on the need for building explanations that the user can understand. However, different approaches on how to get there appear. Different algorithmic approaches are capable of producing different xAI outputs, depending on the level of dependency on the model [21], [33]. The literature is rather divided on the approaches of how explanations are built as scholars either draw on theoretical underpinnings from decision making theories, philosophy and psychology [8], [65], [67], [68] or on very sparse information on human understanding focusing on computational problems, extraction of performance measures and produce transparency in the model output [23], [30], [31], [33], [69]. "*Much of the research about how to interpret and explain AI behavior, they say, is driven by the needs of those who build AI, and not necessarily of those who use it*" [23, p. 3].

On the other hand, studies that take a social perspective focus on how humans experience explanations:

1. Explanations are contrastive: Sought for in response to counterfactual cases (foils). People ask why P happened instead of Q.

2. Explanations are selected - in a biased manner: People do not expect explanations that consist of actual and complete cause of an event.

3. Probabilities in explanations do not matter, causalities do: Referring to probabilities or statistical relationships in an explanation is not as effective as referring to causes.

4. Explanations are social: They are presented in a context through transfer knowledge, interactions, and conversations. People interact differently to explanations [8].

The divide between these two approaches towards building xAI constitutes a research gap from computer science to social science. It also leaves an impression of xAI users being either lay-users or developers with only a few articles discussing the different roles in the processes of xAI.

Few papers address the different AI literacy levels users may have and even fewer address the diversity of stakeholders and their different needs for xAI. While there are various levels of AI literacy and diverse subject domains, researchers focus on developing a user-centric conceptual framework [65]. This seems to be the norm, as the users are rarely defined as anything else as 'user', though sometimes, when case examples are presented, users are mentioned as e.g. clinicians, doctors or experts [54], [55], [65], [70]–[73]. Only a few papers discuss different types of roles and stakeholders in regard to the ecosystem of xAI, and argue that one solution might not fit the purpose of all different types of users but we need to include the context, background and knowledge of the stakeholders to produce understandable explanations [52], [74].

For identifying the different stakeholders Arrieta et al. [22] presents a framework – or target audience, in relation to xAI and acknowledge that stakeholders have different explainability needs in ML models. For example, domain experts may seek out xAI to gain scientific knowledge, whereas users affected by the model's decision may seek to better understand and verify a fair decision was made. Arrieta et al. [22] continue to present a taxonomy of the different techniques that produce xAI within images, text and tabular data.

### 3.4. The technologies producing xAI

Many different approaches towards building more transparent and explainable models have been presented in recent years in the search for opening the infamous black-box. Models such as PD plots (Partial Dependencies), ALE plots (Accumulated Local Effects) ICE (Individual Conditional Expectation), SHAP values (SHapely Additive exPlanations) and LIME (Local Interpretation) are amongst the most popular groundworks for feature-based models, excluding image-based algorithms such as CNN [30]–[33], [40], [75]–[78]. The above mentioned methods each have their different approaches but can be classified as follows:

**Intrinsically transparent**: ML Models that are of a simpler character, but also less precise than other more advanced models (Linear Regression, Logistic Regression, Decision Trees RuleFit) [22], [32].

**Model agnostic xAI frameworks**: These are often of a post-hoc character, meaning that they are designed to fit any model type and rely on techniques that simplifies the model (rule-based extraction from the complex model), show feature relevance estimations (e.g. SHAP produces an additive feature importance score for single predictions), visualizes the model (e.g. ICE which visualizes the estimated model of supervised ML models) or produce a local surrogate model of the output (e.g. LIME)[22], [32], [33], [79], [80]. A common thread of these frameworks is that they produce some kind of visual output for easier understanding. However, a thorough comprehension of the output of additive models, can still be a challenge to laymen [49], [82], [83].

## 4. Discussion and Future Research Agenda

The analysis of the xAI literature shows that due to the rather new and interdisciplinary field of xAI [24], a common understanding of key concepts has not been fully established yet. However, more recent work [8], [22] is providing more comprehensive definitions having a greater impact and shaping the field.

The nascency of the field is also evident in the approaches towards building and designing xAI frameworks and outputs which were driven by developers. Recently, researchers have come to understand the need for closer interdisciplinary work, as the explanations originally built for the developers – by developers, can serve more purposes than just ensuring sound and reliable models. xAI as a means to minimize biases, ensure social responsibility and fairness is as much about data preparation and model design as it is about proving that it is taken into account. As a tool for developers, xAI can be of great importance to ensure a model is based on causations and not correlations, data is evenly distributed, and features used are relevant. However, the demand of knowledge from programmers, data scientists and computer scientists is increasing as the stack grows. Moreover, to possess the domain knowledge a clinician might have within his field will never be reasonable, however necessary to build sound models. This again calls for more interdisciplinary research in xAI to ensure we can collaborate across expert domains and obtain the expertise from all related fields. This is in line with recent calls from IS researchers for interdisciplinary research into AI phenomena [11], [84].

Considering the application of xAI in AI based systems, the four thematic debates indicate that organizations need to be cautious when choosing to use an xAI method to address a specific need. For example, additive models generating instance-level explanations might be misleading to some [49], which does not suffice for ensuring compliance or justify explanations - but could suffice more general explanations. As xAI is often seen as an enabler of AI, their goals are different from AI/ML. AI is for optimization, augmentation and automation of decision making, utilizing the ever so strong power of machine learning. xAI is for exploring and understanding the decision made by AI and as a means to validate models and the performance of ML-models. However, validating explanations and xAI outputs is a much more complex process as it includes the perception of the stakeholder who might not possess the technical literacy level required to follow the chain of logic in the model and explanation combined [66], [25]. Further, xAI should be aligned with the decision objective, which should guide the choice of the ML-model, variable selection, hyper parameter settings and other adjustments of the ML-model. Besides the fact that xAI is costly in terms of computational power and time, this in turn adds to the complexity of any AI-solution, which argues for re-evaluating the need for xAI in a solution.

To generate xAI models that satisfy regulators and ensures compliance, especially within healthcare and finance, has also shown to be of high interest as this is a great entry-barrier for implementing ML in these highly regulated areas [21], [22], [85], [86]. The literature greatly suggests that xAI will ease this barrier, however, there has not yet been any significant empirical studies showing how xAI in fact may satisfy regulators need to ensure compliance or undergo audits. Therefore, we greatly encourage researchers to address this issue with more empirical studies such as Lauritsen et al. [87].

Another emerging topic in xAI is how to validate the explanations and related frameworks. On one hand, researchers need to ensure the human interpretation is accommodated and on the other hand, that the frameworks show information as it is, without skewing the measures and building persuasive outputs instead. We found that different streams of research seem to favor either building intrinsically transparent models, or post-hoc local explanations. Future research should focus on meeting both these contradictory yet interrelated needs combining process transparency and storytelling of the outcome [88], [89].

Moreover, reviewing the demand for knowledge and how much xAI stakeholders need to understand in terms of ensuring reliability in using it, was discussed as there seems to be only little research into how much information we need to interpret and understand. The debate of completeness versus interpretability will not disappear but the nature of users' expertise will influence the level of sophistication in the explanation [25]. However, empirical studies are currently needed to further investigate the relationship between completeness and interpretability, but also if less interpretability could possibly satisfy stakeholders needs without becoming suggestive [91].

The six sub-themes which emerged throughout the literature review constituting the motivation for xAI, are highly interlinked, but each has their own underlying purpose. We argue for further research into building – or defining existing methods that fits the purpose of the motivated need for xAI. This could be either ensuring compliance or minimizing biases in model performance. Developing and implementing ML and xAI we need to ensure the chain of logic persists both in development and production, and that the chosen xAI framework fits the ML model and supports the underlying goal of the intended explanation. One could argue that the same methods might be feasible in many cases, however, the goals of the underlying ML models, and the explanations themselves are different. Therefore, we call for further research in understanding what kinds of explanation techniques fit the different purposes intended.

The thematic debate of human explanations is currently dominated by social science perspectives with little attention to what the technology is capable of producing. This reinforces the social - technical dichotomy making for a very divided xAI field: focusing either on how we as humans interpret explanations or how we can technically extract information from complex models. While the current state is justified by the nascent stage the field is in, further research is required in understanding the different xAI stakeholders needs and how they can be satisfied with building more targeted explanations than the two dominant groups of either developers or users.

Techniques and frameworks developed to produce explanations are continuously evolving, including more and more complex ML models. However, these techniques and frameworks seldomly consider other stakeholders besides developers themselves. Therefore, continuing the path paved by Ribeiro et al. [33] we call for further empirical research on how well the frameworks are understood by different stakeholders. We further argue for research in understanding what xAI frameworks would support each purpose the best, and which would be inexpedient.

While this research agenda is not exhaustive in any way, it revealed two future avenues for xAI research:

a) the need to account for the different stakeholders and their different explanation needs. This research avenue highlights the importance of investigating the micropolitics of xAI in organizations and its implications for work.

b) the need for a holistic approach in investigating xAI, jointly considering the social and technical aspects of xAI, the process and the outcome aspects of xAI, as well as the factual and the storytelling aspects of xAI. This research avenue emphasizes the need to understand conceptually and empirically the complex nature of xAI as a new kind of sociotechnical system and its implications for AI practices in business and society.

## 5. Conclusion

The quest for opening the famous 'AI black box' has attained great traction in the past decade, as ethical concerns, regulations, and the need for controlling these models has increased. Responding to the calls by Miller [8], and Lipton [9] to generate further perspectives and critical writings on xAI we perform a systematic literature review on the need for xAI. We first identify four thematic debates central to how xAI addresses the need for explainable AI. Second, taking a sociotechnical perspective in assessing the debates we identify two future research avenues: a) the need for a stakeholder approach and, b) the need for a holistic view on xAI. Third, we argue that to advance theories and practice on xAI, the IS field is in need of empirical studies that show how different xAI frameworks address the different stakeholder needs. Based on future empirical evidence, we as scholars may be able to judge to what extent xAI meets its expectations, both for its stakeholders seeking to strategically benefit from AI, and society as a whole.